\documentclass[11pt,a4paper]{article}
\pdfoutput=1
\usepackage{jheppub}
\usepackage{tikz}
\usetikzlibrary{intersections, calc, arrows.meta, bending}
\usepackage{subcaption}
\usepackage{simpler-wick}

\DeclareMathOperator{\Tr}{Tr}
\DeclareMathOperator{\tr}{tr}

\newcommand{\ri}{\mathrm{i}}
\renewcommand{\th}{\theta}

\newcommand{\cob}{\delta}

\newcommand{\hf}{\frac{1}{2}}
\newcommand{\qu}{\frac{1}{4}}
\newcommand{\til}[1]{\widetilde{#1}}

\renewcommand{\b}[1]{\overline{#1}}

\newcommand{\bra}{\langle}
\newcommand{\ket}{\rangle}
\newcommand{\la}{\lambda}

\newcommand{\ka}{\kappa}

\newcommand{\bt}{\beta}

\newcommand{\al}{\alpha}
\newcommand{\om}{\omega}

\newcommand{\rt}[1]{\sqrt{#1}}
\newcommand{\cO}{\mathcal{O}}
\newcommand{\cZ}{\mathcal{Z}}

\newcommand{\cN}{\mathcal{N}}
\newcommand{\cM}{\mathcal{M}}

\makeatletter
\gdef\@fpheader{}
\makeatother
\begin{document}
\title{Discrete analogue of the Weil-Petersson volume in double scaled SYK}

\author{Kazumi Okuyama}

\affiliation{Department of Physics, 
Shinshu University, 3-1-1 Asahi, Matsumoto 390-8621, Japan}

\emailAdd{kazumi@azusa.shinshu-u.ac.jp}

\abstract{We show that the connected correlators of partition functions
in double scaled SYK model can be decomposed into ``trumpet'' and 
the discrete analogue of the Weil-Petersson volume, which was defined by Norbury and Scott. We explicitly compute this discrete volume for the first few orders in the genus expansion
and confirm that the discrete volume reduces to the Weil-Petersson volume in a certain semi-classical limit.}

\maketitle

\section{Introduction}
The Sachdev-Ye-Kitaev (SYK) model is a very useful toy model for the
study of quantum gravity  
\cite{Sachdev1993,Kitaev1,Kitaev2,Polchinski:2016xgd,Maldacena:2016hyu}.
At low energy, the SYK model is described by the Schwarzian mode
which is holographically dual to the Jackiw-Teitelboim (JT) gravity
\cite{Jackiw:1984je,Teitelboim:1983ux}.
In the seminal paper \cite{Saad:2019lba}, 
it was found that JT gravity is equivalent to a random matrix model
in the double scaling limit
and the higher genus amplitudes of JT gravity are obtained by
gluing the Weil-Petersson volume
with the so-called trumpet partition functions
and integrating over the lengths of the geodesic loops.

One can go beyond the realm of the low energy Schwarzian
approximation by taking a certain double scaling limit
of the SYK model \cite{Cotler:2016fpe,Berkooz:2018jqr}, 
which we will call DSSYK in this paper.
As shown in \cite{Berkooz:2018jqr}, DSSYK is exactly solvable in the planar limit 
using the technique of the chord diagram and the transfer matrix.
In \cite{Lin:2022rbf} it is suggested that
the chord number state $|n\ket$ appearing in the transfer matrix formalism
can be thought of as a state on the bulk Hilbert space
and the bulk geodesic length is replaced by a discrete chord number $n$
in DSSYK. In \cite{Okuyama:2023byh},
the amplitude of a half-wormhole ending on the end of the world brane 
was computed and it was found that
the length of geodesic loop 
is also discretized in units of the coupling $\la$ of DSSYK. 

So far, the non-planar contributions of DSSYK
have not been fully explored in the literature before.\footnote{See \cite{Berkooz:2020fvm} 
for a work in this direction.}
In this paper, we will study the $1/N$ expansion of the matrix model of DSSYK introduced in \cite{Jafferis:2022wez}.
We focus on the correlator of thermal partition functions of DSSYK
and ignore the effect of matter field for simplicity.
In this case, DSSYK is described by a hermitian one-matrix model
with a certain potential $V(M)$, whose explicit form was obtained in 
\cite{Jafferis:2022wez}.
We find that $n$-point function of the partition function $Z(\bt)$
at genus-$g$ is expanded as
\begin{equation}
\begin{aligned}
 Z_{g,n}(\bt_1,\cdots,\bt_n)=\sum_{b_1,\cdots,b_n\in\mathbb{Z}_{+}}
N_{g,n}(b_1,\cdots,b_n)\prod_{i=1}^n b_iZ_{\text{trumpet}}(\bt_i,b_i),
\end{aligned} 
\label{eq:Zgn-intro}
\end{equation}
where $N_{g,n}$ is a discrete analogue of the volume of 
moduli space of Riemann surfaces introduced by Norbury and Scott \cite{norbury2013polynomials}.
It turns out that the trumpet partition function $Z_{\text{trumpet}}(\bt,b)$
is given by the modified Bessel function
of the first kind, which agrees with the result obtained from the analysis
of the end of the world brane \cite{Okuyama:2023byh}.
We explicitly compute the discrete volume $N_{g,n}$ for small $(g,n)$
in the 
matrix model of DSSYK using the technique of Eynard-Orantin's 
topological recursion \cite{Eynard:2007kz}.
We find that the discrete volume $N_{g,n}$
of DSSYK reduces to the Weil-Petersson volume in a certain semi-classical limit.

This paper is organized as follows.
In section \ref{sec:matrix}, we review the matrix model 
description of DSSYK obtained in \cite{Jafferis:2022wez}.
In section \ref{sec:curve},
we introduce the Joukowsky map and compute the spectral curve 
of the matrix model of DSSYK in \cite{Jafferis:2022wez}.
In section \ref{sec:top-rec}, we review the 
Eynard-Orantin's topological recursion and the discrete volume
$N_{g,n}$ introduced by Norbury and Scott \cite{norbury2013polynomials}. We show that the connected correlator of partition functions
can be decomposed into the discrete volume and the trumpet.
We find that the trumpet partition function is given by the modified Bessel function
of the first kind. 
In section \ref{sec:gauss}, we compute 
the discrete volumes $N_{g,n}$ for the Gaussian matrix model via
the topological recursion and check that they are 
consistent with the known results of the correlators of Gaussian matrix model.  
In section \ref{sec:dssyk}, we compute 
the discrete volumes $N_{g,n}$ with small $(g,n)$ for the matrix model of DSSYK
and confirm that they reduce to the Weil-Petersson volume
in the semi-classical limit.
Finally, we conclude in section \ref{sec:conclusion}
with some discussion on future problems.

\section{Matrix model of DSSYK}\label{sec:matrix}
We first review the result of DSSYK \cite{Berkooz:2018jqr}
and its matrix model representation
\cite{Jafferis:2022wez}. In this paper, we will consider the 
one-matrix model associated with the random Hamiltonian 
of the SYK model and we will ignore the effect of matter operators
for simplicity. 

SYK model is defined by the Hamiltonian for 
$N$ Majorana fermions $\psi_i~(i=1,\cdots,N)$
obeying $\{\psi_i,\psi_j\}=2\cob_{i,j}$
with all-to-all $p$-body interaction
\begin{equation}
\begin{aligned}
 H=\ri^{p/2}\sum_{1\leq i_1<\cdots<i_p\leq N}
J_{i_1\cdots i_p}\psi_{i_1}\cdots\psi_{i_p},
\end{aligned} 
\end{equation}
where $J_{i_1\cdots i_p}$ is a random coupling drawn from the Gaussian distribution
with the mean and the variance given by
\begin{equation}
\begin{aligned}
 \bra J_{i_1\cdots i_p}\ket_J=0,\quad \bra J_{i_1\cdots i_p}^2\ket_J=\binom{N}{p}^{-1}.
\end{aligned} 
\end{equation}
Here and in what follows, $\bra\cdots\ket_J$ refers to the ensemble average
over the random coupling $J_{i_1\cdots i_p}$.

DSSYK is defined by the scaling limit
\begin{equation}
\begin{aligned}
 N,p\to\infty\quad\text{with}\quad\la=\frac{2p^2}{N}:\text{fixed}.
\end{aligned} 
\label{eq:scaling}
\end{equation}
As shown in \cite{Berkooz:2018jqr}, the ensemble average of the moment $\tr H^k$ 
reduces to a counting problem
of the intersection number of chord diagrams
\begin{equation}
\begin{aligned}
 \bra \tr H^k\ket_J=\sum_{\text{chord diagrams}}q^{\#(\text{intersections})},
\end{aligned} 
\end{equation}
where $q$ is given by
\begin{equation}
\begin{aligned}
 q=e^{-\la}.
\end{aligned} 
\end{equation}
Here, $\tr$ in $\tr H^k$ refers to the trace over the Fock space of Majorana fermions.
Using the technique of the transfer matrix,
the disk amplitude $\bra \tr e^{-\bt H}\ket_J$ of DSSYK 
is explicitly evaluated as \cite{Berkooz:2018jqr}
\begin{equation}
\begin{aligned}
 \bra \tr e^{-\bt H}\ket_J=\int_0^\pi\frac{d\th}{2\pi}
(q,e^{\pm2\ri\th};q)_\infty e^{-\bt E(\th)},
\end{aligned} 
\label{eq:disk-DSSYK}
\end{equation}
where $E(\th)$ is given by
\begin{equation}
\begin{aligned}
 E(\th)=-a\cos\th,\qquad a=\frac{2}{\rt{1-q}}.
\end{aligned} 
\label{eq:E-th}
\end{equation}
The $q$-Pochhammer symbol is defined by
\begin{equation}
\begin{aligned}
 (a;q)_\infty=\prod_{k=0}^\infty (1-aq^k),
\end{aligned} 
\end{equation}
and the measure factor in \eqref{eq:disk-DSSYK}
is a shorthand of
\begin{equation}
\begin{aligned}
 (q,e^{\pm2\ri\th};q)_\infty&=(q;q)_\infty (e^{2\ri\th};q)_\infty
(e^{-2\ri\th};q)_\infty.
\end{aligned} 
\end{equation} 

As argued in \cite{Jafferis:2022wez}, one can introduce an
$N\times N$ hermitian one-matrix model
which reproduces \eqref{eq:disk-DSSYK} at genus-zero in the $1/N$ expansion.
We define the partition function of matrix model by
\begin{equation}
\begin{aligned}
 \cZ=\int dM e^{-N\Tr V(M)}
\end{aligned} 
\end{equation}
and the correlator of $Z(\bt)=\Tr e^{\bt M}$ is written as
\begin{equation}
\begin{aligned}
 \Biggl\bra\prod_{i=1}^nZ(\bt_i) \Biggr\ket&=\frac{1}{\cZ}
\int dM e^{-N\Tr V(M)}\prod_{i=1}^n \Tr e^{\bt_i M}.
\end{aligned} 
\end{equation}
In what follows, we will assume that $V(M)$ is an even function of $M$
\begin{equation}
\begin{aligned}
 V(-M)=V(M),
\end{aligned} 
\end{equation}
and the eigenvalues are distributed along the segment
$x\in[-a,a]$ on the real $x$-axis in the large $N$ limit, 
where $x$ denotes the eigenvalue of 
the random matrix $M$.

As usual, the connected correlator of $Z(\bt)$'s admits
the genus expansion in the large $N$ limit
\begin{equation}
\begin{aligned}
 \Biggl\bra\prod_{i=1}^nZ(\bt_i) \Biggr\ket_{\text{conn}}
=\sum_{g=0}^\infty N^{2-2g-n}Z_{g,n}(\bt_1,\cdots,\bt_n).
\end{aligned} 
\label{eq:Zgn}
\end{equation}
Also, the connected correlator
of $\tr e^{-\bt H}$ of DSSYK has a similar large $N$ expansion
\begin{equation}
\begin{aligned}
 \left\bra\prod_{i=1}^n\tr e^{-\bt_i H}\right\ket_{J}^{\text{conn}}
=\sum_{g=0}^\infty N^{2-2g-n}Z_{g,n}'(\bt_1,\cdots,\bt_n).
\end{aligned} 
\label{eq:Z'gn}
\end{equation}
As argued in \cite{Jafferis:2022wez}, the leading order (i.e. genus-zero)
contribution for the $n$-point function of DSSYK
$Z_{0,n}'(\bt_1,\cdots,\bt_n)$ in \eqref{eq:Z'gn} agrees with the matrix model result $Z_{0,n}(\bt_1,\cdots,\bt_n)$ 
in \eqref{eq:Zgn} by 
construction of the matrix model potential in \cite{Jafferis:2022wez}. 
In particular, the genus-zero one-point function $Z_{0,1}(\bt)$
of the matrix model is equal to
the disk amplitude $\bra \tr e^{-\bt H}\ket_J$ of DSSYK 
in \eqref{eq:disk-DSSYK}.
In terms of the genus-zero eigenvalue density $\rho_0(x)$ of the matrix model,
the disk amplitude is written as
\begin{equation}
\begin{aligned}
 \bra \tr e^{-\bt H}\ket_J=Z_{0,1}(\bt)=
\int_{-a}^a dx\,\rho_0(x) e^{\bt x}.
\end{aligned} 
\end{equation}
By identifying $x$ with $-E(\th)$ in \eqref{eq:E-th}
\begin{equation}
\begin{aligned}
 x=-E(\th)=a\cos\th,
\end{aligned} 
\end{equation}
we can read off $\rho_0(x)$ from \eqref{eq:disk-DSSYK}
\begin{equation}
\begin{aligned}
 \rho_0(x)&=\frac{1}{2\pi\rt{a^2-x^2}}(q,e^{\pm2\ri\th};q)_\infty\\
&=\frac{1}{2\pi\rt{a^2-x^2}}\sum_{n\in\mathbb{Z}}
(-1)^nq^{\hf n^2}\bigl[q^{\hf n}+q^{-\hf n}\bigr]T_{2n}(a^{-1}x),
\end{aligned} 
\end{equation}
where $T_n(\cos\th)=\cos(n\th)$ denotes the Chebyshev polynomial of the first kind.

\begin{figure}[t]
\centering
\subcaptionbox{$q=0.1$\label{sfig:q1}}{\includegraphics
[width=0.45\linewidth]{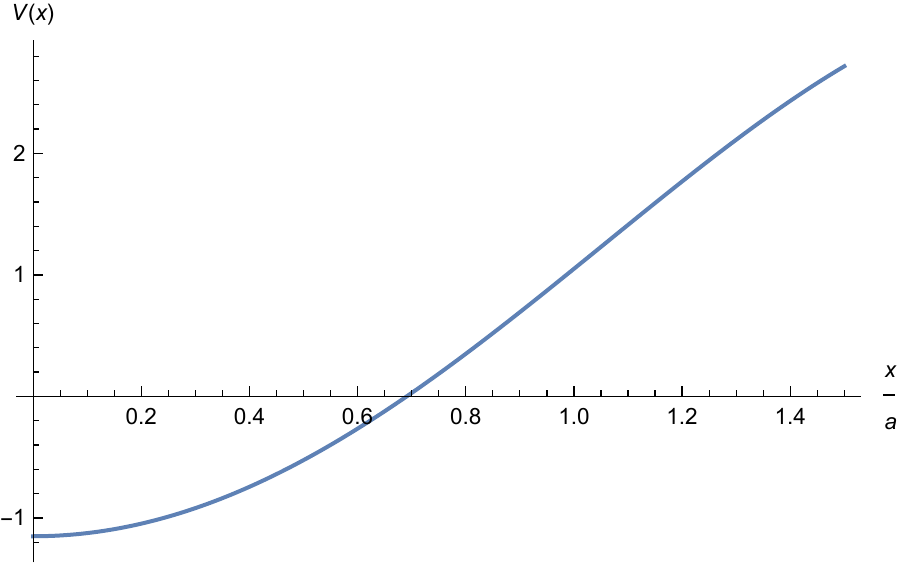}}
\hskip5mm
\subcaptionbox{$q=0.9$\label{sfig:q9}}{\includegraphics
[width=0.45\linewidth]{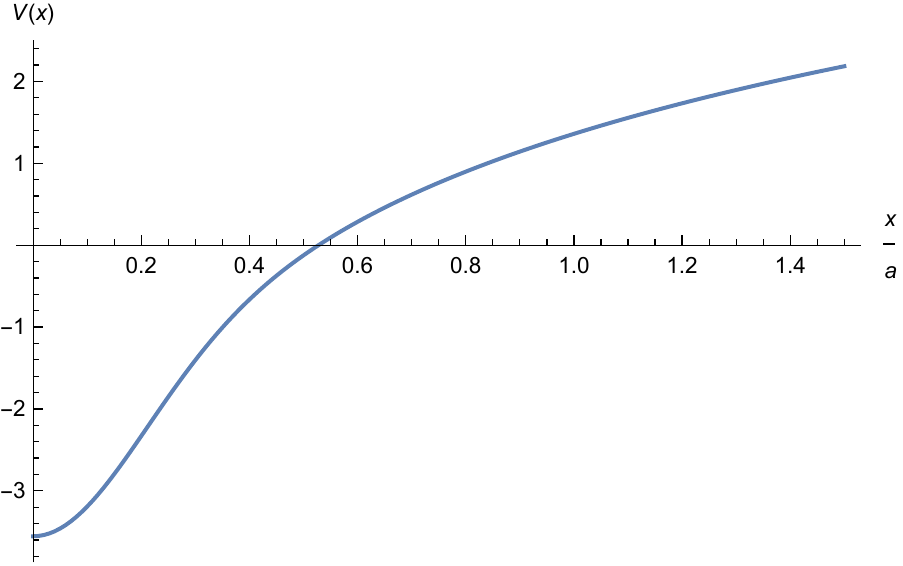}}
  \caption{
Plot 
of the matrix model potential $V(x)$ in \eqref{eq:V-DSSYK} as a function of $x/a$
with $a=\frac{2}{\rt{1-q}}$. We show the plot of $V(x)$ for \subref{sfig:q1} $q=0.1$ and 
\subref{sfig:q9} $q=0.9$.
}
  \label{fig:V}
\end{figure}

From the relation between the eigenvalue density $\rho_0(x)$
and the matrix model potential $V(M)$
\begin{equation}
\begin{aligned}
 \hf V'(x)=\int_{-a}^a dx' \frac{\rho_0(x')}{x-x'},\quad (x\in[-a,a]),
\end{aligned} 
\label{eq:rho-V}
\end{equation}
we find
\begin{equation}
\begin{aligned}
 \hf V'(x)=\frac{1}{a}\sum_{n=1}^\infty
(-1)^{n-1}
q^{\hf n^2}\bigl[q^{\hf n}+q^{-\hf n}\bigr]U_{2n-1}(a^{-1}x),
\end{aligned} 
\label{eq:V'}
\end{equation}
where $U_n(\cos\th)=\frac{\sin(n+1)\th}{\sin\th}$ denotes the 
Chebyshev polynomial of the second kind. 
After integrating $V'(x)$ in \eqref{eq:V'}, the matrix model potential 
$V(M)$ of DSSYK is found to be \cite{Jafferis:2022wez}
\begin{equation}
\begin{aligned}
 V(x)=\sum_{n=1}^\infty
\frac{(-1)^{n-1}}{n}
q^{\hf n^2}\bigl[q^{\hf n}+q^{-\hf n}\bigr]T_{2n}(a^{-1}x).
\end{aligned} 
\label{eq:V-DSSYK}
\end{equation}
Note that in the limit $q\to0$ the potential becomes Gaussian
\begin{equation}
\begin{aligned}
 \lim_{q\to0}V(x)=T_2(x/2)=\hf x^2+\text{const}.
\end{aligned} 
\end{equation}
In Figure \ref{fig:V} we show the plot of $V(x)$ 
in \eqref{eq:V-DSSYK} for $q=0.1$ and $q=0.9$.

For general $g\geq1$, there is no obvious reason to expect the agreement of 
the matrix model result $Z_{g,n}$ in \eqref{eq:Zgn} and
the large $N$ expansion of DSSYK $Z'_{g,n}$ in \eqref{eq:Z'gn}.
Nevertheless, it is tempting to speculate that these two computations actually
agree
at all genera
\begin{equation}
\begin{aligned}
 Z_{g,n}(\bt_1,\cdots,\bt_n)=Z_{g,n}'(\bt_1,\cdots,\bt_n)\quad \forall g\geq0.
\end{aligned} 
\label{eq:conjecture}
\end{equation}
This conjecture \eqref{eq:conjecture} is based on a possible 
holographic dual of DSSYK proposed in \cite{Blommaert:2023opb}.
It is proposed in \cite{Blommaert:2023opb} that, if we ignore the effect of bulk matter fields,
the bulk dual of DSSYK is described as a Poisson-sigma model 
whose bulk dynamics is topological in nature, in a similar manner as
the BF theory description of JT gravity \cite{Saad:2019lba}.
Thus we expect that the path integral of the bulk theory of DSSYK
reduces to a computation of a certain volume of the moduli space of Riemann surfaces,
which in our case corresponds to the discrete volume $N_{g,n}$.
As shown in \cite{norbury2013polynomials}, $N_{g,n}$
obeys the topological recursion and $N_{g,n}$ at the higher genus $g\geq1$
is uniquely determined by the genus-zero data $N_{0,n'}$.
This suggests that the agreement of the matrix model result $Z_{0,n}$
and the DSSYK computation $Z_{0,n}'$ at genus-zero implies the
agreement $Z_{g,n}=Z'_{g,n}$
at higher genus $g\geq1$ assuming that the bulk dual of DSSYK becomes a topological theory
when we ignore the effect of bulk matter fields.
Making this argument as a more rigorous
proof of \eqref{eq:conjecture} is beyond the scope of this paper.
In the rest of this paper, we will concentrate on the higher genus computation
on the matrix model side \eqref{eq:Zgn}, 
which is of interest in its own right regardless of the validity
of the conjecture \eqref{eq:conjecture}.

\section{Joukowsky map and spectral curve}\label{sec:curve}
When the eigenvalues of random matrix 
$M$ are distributed along the segment $x\in[-a,a]$,
we say that such a matrix model is in the one-cut phase.
In this case, the spectral curve of the matrix model
is a Riemann surface of genus-zero and
the square-root branch-cut can be uniformized by the 
Joukowsky map (see e.g. \cite{Eynard:2008we} for a review).

It is well-known that the genus-zero resolvent
\begin{equation}
\begin{aligned}
 R(x)=\left\bra\frac{1}{N}\Tr\frac{1}{x-M}\right\ket_{g=0}
\end{aligned} 
\end{equation}
satisfies the loop equation
\begin{equation}
\begin{aligned}
 R(x)^2=V'(x)R(x)-P(x)
\end{aligned} 
\label{eq:R-rel}
\end{equation}
with
\begin{equation}
\begin{aligned}
 P(x)=\left\bra\frac{1}{N}\Tr\frac{V'(x)-V'(M)}{x-M}\right\ket_{g=0}.
\end{aligned} 
\end{equation}
The loop equation \eqref{eq:R-rel} can be written as
a spectral curve for the matrix model
\begin{equation}
\begin{aligned}
 y^2=\qu V'(x)^2-P(x),
\end{aligned} 
\label{eq:spec-curve}
\end{equation}
where $y$ is defined by
\begin{equation}
\begin{aligned}
 y=\hf V'(x)-R(x).
\end{aligned} 
\label{eq:def-y}
\end{equation}
In the one-cut phase with the support of eigenvalues along the
segment $x\in[-a,a]$, $y$ takes the form
\begin{equation}
\begin{aligned}
 y=Q(x)\rt{x^2-a^2},
\end{aligned} 
\label{eq:y-root}
\end{equation}
where $Q(x)$ is a regular function of $x$.
In other words, $y$ has a square-root branch-point at $x=\pm a$.  
In this case, the spectral curve \eqref{eq:spec-curve} 
can be uniformized by introducing the uniformization
parameter $z$
\begin{equation}
\begin{aligned}
 x(z)=\frac{a}{2}(z+z^{-1}).
\end{aligned}
\label{eq:x-z} 
\end{equation}
This is known as the Joukowsky map. 
Then $y$ in \eqref{eq:y-root} becomes
\begin{equation}
\begin{aligned}
 y=Q\big(x(z)\big)\cdot\frac{a}{2}(z-z^{-1}).
\end{aligned}
\label{eq:y-z} 
\end{equation}
One can see that 
the two branches of the square-root $\rt{x^2-a^2}=\frac{a}{2}(z-z^{-1})$
are exchanged by sending $z$ to its conjugated point $\b{z}=z^{-1}$
\begin{equation}
\begin{aligned}
 z\to\b{z}=z^{-1}~~~\Rightarrow~~~y\to -y.
\end{aligned} 
\end{equation}  
Note that $x(z)=x(\b{z})$ and the cut $[-a,a]$ on the $x$-plane
is mapped to the unit circle on the $z$-plane
\begin{equation}
\begin{aligned}
 x=a\cos\th~~~\Rightarrow~~~z=e^{\ri\th}.
\end{aligned} 
\end{equation} 
In particular, the branch-points $x=\pm a$ correspond to $z=\pm1$.
Also note that the first and the second sheet of the square-root on the $x$-plane
correspond to the inside and the outside of the unit circle on the $z$-plane.

The Joukowsky map enables us to solve the genus-zero loop equation
\eqref{eq:R-rel} by an algebraic means.
From \eqref{eq:def-y} and \eqref{eq:y-z}, one can show that
\begin{equation}
\begin{aligned}
 V'\big(x(z)\big)=R(z)+R(\b{z}),
\end{aligned} 
\label{eq:VR-eq}
\end{equation} 
where we defined $R(z)=R\big(x(z)\big)$. This relation has been already appeared in 
\eqref{eq:rho-V}. Expanding the left hand side of \eqref{eq:VR-eq} as
\begin{equation}
\begin{aligned}
 V'\big(x(z)\big)=\sum_{k\geq1} u_k(z^k+z^{-k})
\end{aligned} 
\label{eq:V-z}
\end{equation}
and using the large $x$ asymptotics of $R(x)\sim 1/x+\cO(x^{-2})$, we find
\begin{equation}
\begin{aligned}
 R(z)=\sum_{k\geq1} u_kz^{-k}.
\end{aligned} 
\label{eq:R-z}
\end{equation}
Plugging \eqref{eq:V-z} and \eqref{eq:R-z} into \eqref{eq:def-y},
$y$ is given by
\begin{equation}
\begin{aligned}
 y=\hf \sum_{k\geq1}u_k(z^k-z^{-k}).
\end{aligned} 
\label{eq:y-u}
\end{equation}
This $y(z)$ together with $x(z)$ in \eqref{eq:x-z}
defines a spectral curve of the hermitian matrix model
in the one-cut phase.\footnote{Our $y(z)$ and $R(z)$ are related to $y(z)$
in \cite{Eynard:2008we} as
\begin{equation}
\begin{aligned}
 R(z)=-y^{\text{there}}(z),\qquad y^{\text{here}}(z)=\hf y^{\text{there}}(z)
-\hf y^{\text{there}}(\b{z}).
\end{aligned} 
\end{equation}}

\subsection{Spectral curve of DSSYK}
Let us apply the above formalism to the matrix model of DSSYK.
Using the relation
\begin{equation}
\begin{aligned}
 U_{2n-1}(w)=2\sum_{j=0}^{n-1} T_{2j+1}(w),
\end{aligned} 
\end{equation}
$V'(x)$ in \eqref{eq:V'} becomes
\begin{equation}
\begin{aligned}
 V'\bigl(x(z)\bigr)=\frac{2}{a}
\sum_{n=1}^\infty(-1)^{n-1}q^{\hf n^2}\bigl[q^{\hf n}+q^{-\hf n}\bigr]
\sum_{j=0}^{n-1}(z^{2j+1}+z^{-2j-1}).
\end{aligned} 
\end{equation}
From \eqref{eq:V-z} and \eqref{eq:y-u}, we can read off $y(z)$ for the DSSYK as
\begin{equation}
\begin{aligned}
 y(z)&=\frac{1}{a}
\sum_{n=1}^\infty(-1)^{n-1}q^{\hf n^2}\bigl[q^{\hf n}+q^{-\hf n}\bigr]
\sum_{j=0}^{n-1}(z^{2j+1}-z^{-2j-1})\\
&=\frac{1}{a}
\sum_{n=1}^\infty(-1)^{n-1}q^{\hf n^2}\bigl[q^{\hf n}+q^{-\hf n}\bigr]
\frac{z^{2n}+z^{-2n}-2}{z-z^{-1}}\\
&=\frac{1}{a}
\sum_{n\in\mathbb{Z}}(-1)^{n-1}q^{\hf n(n+1)}\frac{z^{2n}+z^{-2n}-2}{z-z^{-1}}\\
&=\frac{1}{a}(z-z^{-1})\prod_{n=1}^\infty
(1-q^n)(1-z^2q^n)(1-z^{-2}q^n).
\end{aligned} 
\end{equation}
In the last step, we have used the Jacobi triple product identity
\begin{equation}
\begin{aligned}
 \sum_{n\in\mathbb{Z}}(-1)^{n-1}q^{\hf n(n+1)}w^n=(w^{-1}-1)
\prod_{n=1}^\infty
(1-q^n)(1-wq^n)(1-w^{-1}q^n).
\end{aligned} 
\end{equation}

To summarize, the spectral curve of DSSYK is given by
\begin{equation}
\begin{aligned}
 x(z)=\frac{a}{2}(z+z^{-1}),\quad
y(z)=\frac{1}{a}(z-z^{-1})\prod_{n=1}^\infty
(1-q^n)(1-z^2q^n)(1-z^{-2}q^n),
\end{aligned} 
\label{eq:y-DSSYK}
\end{equation}
with $a=\frac{2}{\rt{1-q}}$.
When $q=0$, \eqref{eq:y-DSSYK} reduces to the spectral curve of the Gaussian
matrix model
\begin{equation}
\begin{aligned}
 x(z)=z+z^{-1},\quad y(z)=\hf(z-z^{-1}).
\end{aligned} 
\label{eq:spec-gauss}
\end{equation}

\section{Trumpet and volume in the one-matrix model}\label{sec:top-rec}
In this section, we will show that the
$n$-point function of partition functions at genus-$g$
is decomposed into the discrete volume $N_{g,n}$ and the trumpet,
as shown in \eqref{eq:Zgn-intro}.

Let us first review the discrete volume 
$N_{g,n}(b_1,\cdots,b_n)$ 
introduced in \cite{norbury2013polynomials} in the case of the one-matrix model.
To this end, it is convenient to consider the connected correlator
of the resolvent $\Tr\frac{1}{x-M}$ instead of $\Tr e^{\bt M}$
\begin{equation}
\begin{aligned}
 \left\bra\prod_{i=1}^n\Tr\frac{1}{x_i-M}\right\ket_{\text{conn}}
=\sum_{g=0}^\infty N^{2-2g-n}W_{g,n}(x_1,\cdots,x_n).
\end{aligned} 
\end{equation}
It is also useful to introduce $\om_{g,n}$ by
\begin{equation}
\begin{aligned}
 \om_{g,n}(z_1,\cdots,z_n)=W_{g,n}\bigl(x(z_1),\cdots,x(z_n)\bigr)
dx(z_1)\cdots dx(z_n).
\end{aligned} 
\end{equation}
For the Joukowsky map $x(z)$ in \eqref{eq:x-z}, $dx(z)$ is given by
\begin{equation}
\begin{aligned}
 dx(z)=\frac{a}{2}(1-z^{-2})dz.
\end{aligned} 
\end{equation}
$\om_{g,n}$ can be computed 
by the Eynard-Orantin's topological recursion \cite{Eynard:2007kz,Eynard:2008we}
\begin{equation}
\begin{aligned}
 \om_{g,n+1}(z_0,J)=\sum_{\al=\pm1}\underset{z=\al}{\text{Res}}K(z_0,z)
\Bigl[\om_{g-1,n+2}(z,\b{z},J)+\sum_{h=0}^g\sum'_{I\subset J}
\om_{h,1+|I|}(z,I)\om_{g-h,1+n-|I|}(\b{z},J\backslash I)\Bigr]
\end{aligned} 
\label{eq:top-rec}
\end{equation}
with the initial condition
\begin{equation}
\begin{aligned}
\om_{0,2}(z_1,z_2)&=\frac{dz_1dz_2}{(z_1-z_2)^2}.
\end{aligned} 
\end{equation}
In \eqref{eq:top-rec}, the prime in the summation means that
$(h,I)=(0,\emptyset),(g,J)$ are excluded
and the recursion kernel $K(z_0,z)$ is given by \footnote{
The Bergmann kernel $B(z_1,z_2)$ is equal to $\om_{0,2}(z_1,z_2)$
in the case of the one-cut one-matrix model.
}
\begin{equation}
\begin{aligned}
 K(z_0,z)&=-\frac{\int_{z'=\b{z}}^z \om_{0,2}(z_0,z')}{4y(z)dx(z)}
=-\frac{z}{2ay(z)(z_0-z)(z_0-z^{-1})dz}dz_0.
\end{aligned} 
\end{equation}

Following \cite{norbury2013polynomials},
we expand $\om_{g,n}$ as
\begin{equation}
\begin{aligned}
 \om_{g,n}(z_1,\cdots,z_n)=\sum_{b_1,\cdots,b_n\in\mathbb{Z}_{+}}
N_{g,n}(b_1,\cdots,b_n)\prod_{i=1}^n b_iz_i^{b_i-1}dz_i.
\end{aligned} 
\label{eq:om-N}
\end{equation}
As discussed in \cite{norbury2013polynomials},
$N_{g,n}(b_1,\cdots,b_n)$ is naturally interpreted as a discrete analogue
of the volume of the moduli space of Riemann surfaces.
In particular, for the Gaussian matrix model
$N_{g,n}(b_1,\cdots,b_n)$ counts the number of lattice points
on the moduli space of curves
\cite{norbury2008counting}.

We can express $Z_{g,n}(\bt_1,\cdots,\bt_n)$ defined in \eqref{eq:Zgn}
in terms of the volume $N_{g,n}(b_1,\cdots,b_n)$ and the discrete version
of the trumpet introduced in \cite{Jafferis:2022wez,Okuyama:2023byh}.
Note that $Z_{g,n}$ is written in terms of 
the genus-$g$ resolvent 
$W_{g,n}$ as
\begin{equation}
\begin{aligned}
 Z_{g,n}(\bt_1,\cdots,\bt_n)=\oint \prod_{i=1}^n \frac{dx_i}{2\pi\ri} e^{\bt_ix_i}
W_{g,n}(x_1,\cdots,x_n),
\end{aligned} 
\label{eq:Z-xint}
\end{equation}
where the contour of integration $\oint dx$ surrounds the cut $[-a,a]$ 
on the $x$-plane.
By the Joukowsky map \eqref{eq:x-z} this contour is mapped to 
a circle surrounding $z=0$ on the $z$-plane
and \eqref{eq:Z-xint} is rewritten as
\begin{equation}
\begin{aligned}
 Z_{g,n}(\bt_1,\cdots,\bt_n)&=\prod_{i=1}^n\oint_{z_i=0}\frac{1}{2\pi\ri}
e^{\frac{\bt_i a}{2}(z_i+z_i^{-1})}\om_{g,n}(z_1,\cdots,z_n).
\end{aligned}
\label{eq:Z-zint}
\end{equation}
Using the relation
\begin{equation}
\begin{aligned}
 e^{\frac{\bt a}{2}(z+z^{-1})}=\sum_{n\in\mathbb{Z}}I_n(\bt a)z^{-n},
\end{aligned} 
\end{equation}
and \eqref{eq:om-N}, the $z_i$-integral in \eqref{eq:Z-zint} is evaluated as
\begin{equation}
\begin{aligned}
 Z_{g,n}(\bt_1,\cdots,\bt_n)
&=\sum_{b_1,\cdots,b_n\in\mathbb{Z}_{+}}
N_{g,n}(b_1,\cdots,b_n)\prod_{i=1}^n b_i
Z_{\text{trumpet}}(\bt_i,b_i),
\end{aligned} 
\label{eq:Z-Ntr}
\end{equation}
where the trumpet partition function is given by
\begin{equation}
\begin{aligned}
 Z_{\text{trumpet}}(\bt,b)=I_b(\bt a).
\end{aligned} 
\label{eq:trumpet}
\end{equation}
Here $I_\nu(z)$ denotes the modified Bessel function of the first kind.
This result of trumpet \eqref{eq:trumpet} 
agrees with the one obtained from the analysis of end of the world brane in 
\cite{Okuyama:2023byh}.

We should stress that the decomposition \eqref{eq:Z-Ntr} of $Z_{g,n}$ 
into the volume $N_{g,n}$ and the trumpet is valid for 
arbitrary one-matrix models in the one-cut phase.
Note also that $Z_{\text{trumpet}}(\bt,b)$ in \eqref{eq:trumpet}
is universal in the sense that it does not depend on the details of
the matrix model potential $V(M)$; it only depends on the endpoint $a$
of the cut.

\section{Discrete volume in the Gaussian matrix model}\label{sec:gauss}
As a warm-up, let us consider the
discrete volume $N_{g,n}$ in the Gaussian matrix model.
One can easily compute $\om_{g,n}$ for small $(g,n)$ 
by solving the topological recursion 
\eqref{eq:top-rec} for the
spectral curve of the Gaussian matrix model \eqref{eq:spec-gauss}.
For instance, $\om_{0,3}$ is given by
\begin{equation}
\begin{aligned}
 \om_{0,3}(z_1,z_2,z_3)&=\sum_{\al=\pm1}
\underset{z=\al}{\text{Res}}K(z_1,z)\bigl[
\om_{0,2}(z,z_2)\om_{0,2}(\b{z},z_3)+
\om_{0,2}(z,z_3)\om_{0,2}(\b{z},z_2)\bigr]\\
&=2\frac{(1+z_1z_2+z_2z_3+z_3z_1)(z_1+z_2+z_3+z_1z_2z_3)}{(1-z_1^2)^2(1-z_1^2)^2(1-z_1^2)^2}dz_1dz_2dz_3.
\end{aligned} 
\end{equation}
From the definition of $N_{g,n}$ in \eqref{eq:om-N}, we can extract
$N_{0,3}$ as
\begin{equation}
\begin{aligned}
 N_{0,3}(b_1,b_2,b_3)=P_{b_1+b_2+b_3},
\end{aligned} 
\label{eq:N03-gauss}
\end{equation}
where $P_b$ denotes the projection to even $b$
\begin{equation}
\begin{aligned}
 P_b=\frac{1+(-1)^b}{2}.
\end{aligned} 
\end{equation}
In a similar manner we find
\begin{equation}
\begin{aligned}
 N_{1,1}(b)&=\frac{b^2-4}{48}P_b,\\
N_{2,1}(b)&=\frac{(b^2-2^2)(b^2-4^2)(b^2-6^2)(5b^2-32)}{8847360}P_b,\\
N_{1,2}(b_1,b_2)&=\frac{(b_1^2+b_2^2-2)(b_1^2+b_2^2-10)}{384}P_{b_1+b_2}
+\frac{1}{32}P_{b_1}P_{b_2}.
\end{aligned} 
\label{eq:N-gauss}
\end{equation}
This agrees with the result in \cite{norbury2008counting} as expected.
As discussed in \cite{Okuyama:2023byh}, for $(g,n)=(0,2)$ it is natural to define
\begin{equation}
\begin{aligned}
 N_{0,2}(b_1,b_2)=\frac{1}{b_1}\cob_{b_1,b_2},
\end{aligned} 
\end{equation}
which reproduces the known result of cylinder amplitude in the Gaussian
matrix model \cite{Akemann:2001st}.

It turns out that the sum over $b_i$ in \eqref{eq:Z-Ntr}
can be performed in a closed form. For instance,
\begin{equation}
\begin{aligned}
 Z_{1,1}(\bt)&=\sum_{b=1}^\infty N_{1,1}(b)bI_b(2\bt)=\frac{(2\bt)^2I_2(2\bt)}{48},\\
Z_{2,1}(\bt)&=\sum_{b=1}^\infty N_{2,1}(b)bI_b(2\bt)=
\frac{(2\bt)^4 I_4(2\bt)}{1280}+\frac{(2\bt)^5I_5(2\bt)}{9216},
\end{aligned} 
\end{equation}
which reproduces the $1/N$ corrections to the $1/2$ BPS Wilson loop
in $\cN=4$ super Yang-Mills \cite{Drukker:2000rr}, which
is described by the Gaussian matrix model 
due to the supersymmetric localization \cite{Pestun:2007rz}.
One can also check that the known results of $Z_{g,n}$ of the 
Gaussian matrix model in \cite{Akemann:2001st,Okuyama:2018aij}
are reproduced from $N_{g,n}$ in \eqref{eq:N03-gauss} and
\eqref{eq:N-gauss} by summing over $b_i$ in \eqref{eq:Z-Ntr}.

As shown in \cite{norbury2008counting},
the semi-classical limit of $N_{g,n}$ reproduces the so-called Kontsevich volume
\begin{equation}
\begin{aligned}
 \lim_{\la\to0}N_{g,n}(\la^{-1}L_1,\cdots, \la^{-1}L_n)
=2^{3-2g-n}\la^{-2(3g-3+n)}V^{\text{K}}(L_1,\cdots,L_n),
\end{aligned} 
\label{eq:limitK}
\end{equation}
where the Kontsevich volume is defined by
\begin{equation}
\begin{aligned}
 V^{\text{K}}(L_1,\cdots,L_n)=\int_{\b{\cM}_{g,n}}\exp\left(\hf\sum_{i=1}^n
L_i^2\psi_i\right).
\end{aligned} 
\label{eq:VK}
\end{equation}
In \eqref{eq:limitK} we assumed $\la^{-1}L_i\in 2\mathbb{Z}_{+}$
and $\psi_i$ in \eqref{eq:VK} is the so-called $\psi$-class on the moduli
space $\b{\cM}_{g,n}$ of genus-$g$ Riemann surfaces
with $n$ punctures.  The power of $\la$ on the right hand side
of \eqref{eq:limitK} is related to the dimension of
$\b{\cM}_{g,n}$
\begin{equation}
\begin{aligned}
 \dim \b{\cM}_{g,n}=2(3g-3+n).
\end{aligned} 
\end{equation}

\section{Discrete volume in DSSYK}\label{sec:dssyk}
Now let us consider the discrete volume $N_{g,n}$ for the matrix model of
DSSYK.
Using the spectral curve of DSSYK in \eqref{eq:y-DSSYK}, we can compute $\om_{g,n}$
by the topological recursion and extract the volume $N_{g,n}$
from the small $z_i$ expansion of $\om_{g,n}$ in \eqref{eq:om-N}.
In this manner, we find the first few terms of $N_{g,n}$
in the matrix model of DSSYK
\begin{equation}
\begin{aligned}
 \til{N}_{0,3}(b_1,b_2,b_3)=&P_{b_1+b_2+b_3},\\
\til{N}_{1,1}(b)=&\left[\frac{b^2-4}{48}+\frac{\zeta_q(2)}{2}\right]P_b,\\
\til{N}_{2,1}(b)=&
\Biggl[\frac{(5b^2-32)(b^2-2^2)(b^2-4^2)(b^2-6^2)}{8847360}
+\frac{(29 b^2-108)(b^2-2^2)(b^2-4^2)}{92160}\zeta_q(2)\\
&+ \frac{(29 b^2-48)(b^2-2^2)}{1536}\zeta_q(4)+\frac{(359 b^2-624)(b^2-2^2)}{7680}\zeta_q(2)^2\\
&+\frac{191 b^2+4}{96}\zeta_q(2)^3+\frac{81 b^2+28}{32}\zeta_q(2)\zeta_q(4)
+\frac{7(5b^2+4)}{48}\zeta_q(6)\\
&+
\frac{845}{48}\zeta_q(2)^4+\frac{185}{16}\zeta_q(4)^2
+\frac{399}{8}\zeta_q(2)^2\zeta_q(4)
+\frac{203}{6}\zeta_q(2)\zeta_q(6)+\frac{105}{8}\zeta_q(8)\Biggr]P_b,\\
\til{N}_{1,2}(b_1,b_2)=&
\Biggl[
\frac{(b_1^2+b_2^2-10)(b_1^2+b_2^2-2)}{384} +\frac{b_1^2+b_2^2-2}{4}\zeta_q(2)
+\frac{7\zeta_q(2)^2}{2}+\frac{5\zeta_q(4)}{2}\Biggl]P_{b_1+b_2}\\
&+\frac{1}{32}P_{b_1}P_{b_2}.
\end{aligned} 
\label{eq:N-DSSYK}
\end{equation}
Here we defined
\begin{equation}
\begin{aligned}
 \til{N}_{g,n}(b_1,\cdots,b_n)=(q;q)_\infty^{3(2g-2+n)}N_{g,n}(b_1,\cdots,b_n),
\end{aligned} 
\end{equation}
and $\zeta_q(s)$ is a $q$-analogue of the zeta function
\begin{equation}
\begin{aligned}
 \zeta_q(s)=\sum_{n=1}^\infty \left(q^{-\hf n}-q^{\hf n}\right)^{-s}.
\end{aligned} 
\end{equation}

In the semi-classical limit $\la\to0$, the discrete length $b_i\in\mathbb{Z}_{+}$
becomes a continuous geodesic length $L_i\in\mathbb{R}_{+}$
and they are related by \cite{Okuyama:2023byh}
\begin{equation}
\begin{aligned}
 L_i=\la b_i,
\end{aligned} 
\end{equation} 
where $\la$ is the coupling of DSSYK defined in \eqref{eq:scaling}.
From our result of $N_{g,n}$ of DSSYK in \eqref{eq:N-DSSYK}, we observe that
$N_{g,n}$ reduces to the Weil-Petersson volume in the semi-classical limit
\begin{equation}
\begin{aligned}
 \lim_{\la\to0}N_{g,n}(\la^{-1}L_1,\cdots,\la^{-1}L_n)
=2c^{2-2g-n}\la^{-2(3g-3+n)}V^{\text{WP}}_{g,n}(L_1,\cdots,L_n).
\end{aligned} 
\label{eq:N-WP}
\end{equation}
Here we assumed $\la^{-1}L_i\in2\mathbb{Z}_{+}$ and $c$ is given by
\begin{equation}
\begin{aligned}
 c=2(q;q)_\infty^3,
\end{aligned} 
\end{equation}
and the Weil-Petersson volume is defined by
\begin{equation}
\begin{aligned}
 V^{\text{WP}}_{g,n}(L_1,\cdots,L_n)=\int_{\b{\cM}_{g,n}}\exp\left(2\pi^2\ka
+\hf\sum_{i=1}^nL_i^2\psi_i\right),
\end{aligned} 
\end{equation}
where $\ka$ denotes the $\ka$-class (also known as the 
first Miller-Morita-Mumford class).
For small $(g,n)$ the explicit form of the 
Weil-Petersson volume reads \footnote{See e.g. \cite{do2011moduli}
for a table of $V^{\text{WP}}_{g,n}(L_1,\cdots,L_n)$ with various $(g,n)$.}
\begin{equation}
\begin{aligned}
 V^{\text{WP}}_{0,3}(L_1,L_2,L_3)&=1,\\
 V^{\text{WP}}_{1,1}(L)&=\frac{L^2}{48}+\frac{\pi^2}{12},\\
V^{\text{WP}}_{2,1}(L)&=\frac{L^8}{442368}+\frac{29\pi^2L^6}{138240}+\frac{139\pi^4L^4}{23040}
+\frac{169\pi^6L^2}{2880}+\frac{29\pi^8}{192},\\
V^{\text{WP}}_{1,2}(L_1,L_2)&=\frac{(L_1^2+L_2^2)^2}{192}+\frac{\pi^2(L_1^2+L_2^2)}{12}+\frac{\pi^4}{4}.
\end{aligned} 
\label{eq:V-WP}
\end{equation}
Using the fact that the small $\la$ limit of $\zeta_q(s)$ with $q=e^{-\la}$
reduces to the ordinary zeta function $\zeta(s)$
\begin{equation}
\begin{aligned}
 \lim_{\la\to0}\zeta_q(s)=\la^{-s}\zeta(s),
\end{aligned} 
\end{equation} 
one can see that the factor of $\pi$ in the Weil-Petersson volume
in \eqref{eq:V-WP} is correctly reproduced from
the semi-classical limit of $N_{g,n}$ in \eqref{eq:N-DSSYK}.

Since the power of $c$ in \eqref{eq:N-WP}
is the Euler characteristic $\chi=2-2g-n$ of the genus-$g$ Riemann surface with $n$
punctures, $c$ can be absorbed into the definition of the genus
counting parameter. The overall factor of $2$ in \eqref{eq:N-WP}
comes from the fact that there are two branch-points $z=\pm1$.

In the semi-classical limit, the sum over $b_i$ in \eqref{eq:Z-Ntr}
becomes the integral over $L_i$ and the trumpet in \eqref{eq:trumpet} reduces to that of JT gravity.
As discussed in \cite{Okuyama:2023byh},
the semi-classical limit of trumpet is obtained by 
zooming in on the edge of the spectrum
$E(\th)=-a$. By rescaling $\th=\la k$ and $b=\la^{-1}L$, 
the semi-classical limit of trumpet in \eqref{eq:trumpet} becomes
\begin{equation}
\begin{aligned}
 I_b(\bt a)&=\int_0^{\pi}\frac{d\th}{2\pi}e^{\bt a\cos\th}2\cos(b\th)\\
&\approx \int_0^{\infty}\frac{\la dk}{\pi}e^{\bt a-\hf \bt a\la^2k^2} \cos(Lk)\\
&=\la e^{\bt a}Z_{\text{trumpet}}(\bt\la^{3/2},L)
\end{aligned} 
\label{eq:lim-trumpet}
\end{equation}
where we defined the continuum version of the trumpet as
\begin{equation}
\begin{aligned}
 Z_{\text{trumpet}}(\bt,L)=
\int_0^\infty\frac{dk}{\pi}e^{-\bt k^2} \cos(Lk)=\frac{e^{-\frac{L^2}{4\bt}}}{\rt{4\pi\bt}}.
\end{aligned} 
\end{equation}
Then \eqref{eq:Z-Ntr} becomes
\begin{equation}
\begin{aligned}
 Z_{g,n}(\bt_1,\cdots,\bt_n)=\prod_{i=1}^n\int_0^\infty dL_i L_i 
Z_{\text{trumpet}}(\bt_i,L_i)V^{\text{WP}}_{g,n}(L_1,\cdots,L_n),
\end{aligned} 
\end{equation}
up to an overall normalization.
This reproduces the result of JT gravity matrix model \cite{Saad:2019lba}, 
as expected.
Thus, our expression of $Z_{g,n}$ in \eqref{eq:Z-Ntr} can be thought of as the DSSYK analogue of the
genus expansion of the multi-boundary correlators, where
the Weil-Petersson volume $V^{\text{WP}}_{g,n}$ is replaced by the discrete volume
$N_{g,n}$.

\section{Conclusions and outlook}\label{sec:conclusion}
In this paper, we have studied the matrix model of DSSYK
in \cite{Jafferis:2022wez} by focusing on the correlators
involving the Hamiltonian only.
Then DSSYK is described by a hermitian one-matrix model
with the potential \eqref{eq:V-DSSYK}.
We found the spectral curve \eqref{eq:y-DSSYK}
of this matrix model and computed the discrete volume $N_{g,n}$
using the technique of Eynard-Orantin's topological recursion.
We have checked that our discrete volume $N_{g,n}$ reduces to
the Weil-Petersson volume in the semi-classical limit \eqref{eq:N-WP}.
As shown in \eqref{eq:Z-Ntr}, the multi-boundary correlator $Z_{g,n}$ of DSSYK
can be constructed by gluing the discrete volume $N_{g,n}$ and the trumpets,
and summing over the discrete lengths $b_i$.
We found that the trumpet is given by the modified Bessel function
\eqref{eq:trumpet} universally for arbitrary one-matrix models in the one-cut phase.

There are several interesting open questions.
Although we have checked the relation \eqref{eq:N-WP}
between the Weil-Petersson volume $V^{\text{WP}}_{g,n}$ and the discrete volume 
$N_{g,n}$ of DSSYK for small $(g,n)$, we do not have a general proof of
\eqref{eq:N-WP}. It would be interesting find a proof of 
\eqref{eq:N-WP} along the lines of \cite{norbury2008counting,norbury2013polynomials}.

We observed that the integral over the geodesic length $L$ in JT gravity
is
replaced by the sum over the discrete length $b$
if we do not take a double scaling limit of matrix model and
consider the ordinary large $N$ 't Hooft expansion.
As we saw in \eqref{eq:lim-trumpet}, 
the double scaling limit of matrix model amounts to taking a low energy 
limit of DSSYK. If we go beyond the realm of low energy approximation
and keep the whole spectrum $E(\th)=-a\cos\th$
of DSSYK, we start to see a discrete structure of the bulk spacetime which
is described by the ordinary large $N$ 't Hooft expansion of the matrix model.
Namely, the bulk spacetime of DSSYK is a discrete random surface
generated by the 't Hooft double-line diagram with polygonal faces \cite{tHooft:1973alw,Brezin:1977sv}. 
This is similar in spirit to the work of Gopakumar and collaborators 
\cite{Gopakumar:2011ev,Gopakumar:2012ny,Gopakumar:2022djw}
on the holography of Gaussian matrix model.
In \cite{Gopakumar:2011ev,Gopakumar:2012ny,Gopakumar:2022djw}, 
it is found that the moduli
integral in the 't Hooft expansion of matrix model receives
contributions only from certain discrete points on the moduli space, which is reminiscent of the counting of lattice points on the moduli space of curves
to define $N_{g,n}$ in \cite{norbury2008counting}.
It would be interesting to develop a lattice counting interpretation of
$N_{g,n}$ for the matrix model of DSSYK as well.

On general grounds, we expect that the $1/N$ expansion of the correlator in \eqref{eq:Zgn} is an asymptotic series and it receives a non-perturbative correction
of the order $e^{-NA(q)}$ with some instanton action $A(q)$.
It would be interesting to study the analogue of ZZ branes in
DSSYK via the approach of resurgence as in \cite{Eynard:2023qdr}.

In this paper, we have ignored the effect of matter field for simplicity.
It is argued in \cite{Jafferis:2022wez} that if we include the
matter operators in DSSYK, it is described by a two-matrix model.
It would be interesting to study the two-matrix model in 
\cite{Jafferis:2022wez} by generalizing our work on the one-matrix sector of DSSYK.
We leave this as an interesting future problem.
 
\acknowledgments
The author would like to thank  Bertrand Eynard for correspondence.
This work was supported
in part by JSPS Grant-in-Aid for Transformative Research Areas (A) 
``Extreme Universe'' 21H05187 and JSPS KAKENHI Grant 22K03594.

\bibliography{paper}
\bibliographystyle{utphys}

\end{document}